\documentclass[
    ,final            
  ]
  {aipproc}

\layoutstyle{6x9}

\begin{document}

\title{ Proton/pion ratios and radial flow in pp and
peripheral heavy ion collisions}

\classification{25.75.Dw :
                \texttt{14.40.-n,14.20.-c }}

\author{E. Cuautle and G. Pai\'c }{
  address={Instituto de Ciencias Nucleares, Universidad Nacional
  Aut\'onoma de M\'exico,\\
 Apartado Postal 70-543, M\'exico, Distrito Federal 04510 M\'exico}
}

\begin{abstract}
The production of baryon and  mesons in the RHIC heavy-ion experiments
has received a lot of attention lately. Although not widely known, the
$pp$ data measured concurrently with  heavy ion collisions do not find a
convincing  explanation in  terms  of simple  models.  We present  the
results  of an  afterburner  to Pythia  and  Hijing event  generators,
simulating  radial  flow  which  seems to  qualitatively  explain  the
experimental results when  applied to the pp collision  data from RHIC
at 200 GeV center-of-mass energy.

\end{abstract}

\maketitle


\section{INTRODUCTION}

In heavy ion  collisions at RHIC energies some  phenomena remain up to
now without valid explanation, one of them is the $p/\pi$ ratios which
exhibits a departure from the  same ratio, measured in $pp$ collisions
at    transverse    momenta     between    2.0    and    $\sim    $4.5
GeV/c\cite{Adcox:2003nr}.   Although the coalescence  model \cite{hwa}
has been widely accepted, it  does not provide a satisfactory response
to many  questions.  Beyond the  details of the  hadronization process
which is  still a debated  question; it is  well known that  the $p_t$
spectra of particles  of different masses cannot be  fit with a unique
temperature as would  predict a naive thermal model.   In the nineties
the NA44 collaboration  has put in evidence the  so called {\it radial
flow}\cite{NA44}  in heavy ion  collisions and  it was  interpreted as
consequence  of  multiple  interaction  among the  partons  after  the
collisions of  ions and before the  freeze out of  the created system.
This interpretation also was  confirmed by recent experimental results
from RHIC~\cite{STAR}.\\

The  flow  is  understood  as  reflecting collective  aspects  of  the
interacting medium, depending on the collision energy.
In central collisions between spherical nuclei, the initial
state  is  symmetric  in   azimuth  implying  an  isotropic  azimuthal
distribution of the final  state particles. Consequently, any pressure
gradient will  cause an azimuthally  symmetric collective flow  of the
outgoing particles. This  is what we call radial  flow.\\ The relevant
observable to study  the radial flow is the  transverse momenta of the
particles.   For   each  particle,   the  random  thermal   motion  is
superimposed    onto   the    collective    radial   flow    velocity,
correspondingly,  the  invariant  $p_t$  distribution depends  of  the
temperature  at  freeze  out,  the  particle mass,  and  the  velocity
profile.  The  experimental data on  radial flow at RHIC  indicate two
things:
\begin{itemize}
\item the  temperature lowers with centrality while  the flow velocity
increases~\cite {XXXXX}.
\item even for $pp$ collisions the analysis of the data from STAR yields
  a flow value of $\approx 0.2$ c
\end{itemize} 
This  latter result  coupled to  the fact  that in  $pp$  collisions the
$p/\pi$ ratio could  not be reproduced by Pythia  has encouraged us to
investigate what  would be the result  of the $p/\pi$ ratio  if a flow
component would be incorporated.
 
Consequently, in     a     first     step     we     have     generated
{\it flow-free}  pion and  proton spectra  using: 1) Pythia
~\cite{Pythia63} with  the Popcorn baryon  production mechanism (note
that from  earlier work~\cite{cuautle05} we know  that the differences
in  the  ratio  at 200  GeV  are  not  very  large when  other  baryon
production mechanisms are used).  2) Hijing 1.36.\\
In a second  step a toy model of the radial  flow is incorporated in
the $p_t$ spectra and finally our results are compared with the baryon
to meson ratio from STAR~\cite{ppiratio} in $pp$.

\section{THE RADIAL FLOW AFTER BURNER}

  We are assuming  that a fireball, thermalized, and expanding was
  created in a collision from an event generator.
The  expansion produces  an
additional  momentum  to  the  one  created in  the  collisions.  This
contribution  we  call  momentum  of  the flow  $p_{t,f}$  given  by
$p_{t,f}  =\gamma  m\beta$,  where  $\gamma$ is  the  Lorentz  factor,
$\beta$ is  the profile velocity and  $m$ is the mass  of the particle
under consideration.   In order  to add this  radial component  to the
transverse  momenta produced  by  the generators,  it  is necessary  to
attribute to the momentum $p_t$ of each particle a randomized position
in the transverse plane. Once  this is done, the radial flow component
is added  vectorially.\\

\section{RESULTS}
\begin{figure}
\includegraphics[height=.3\textheight]{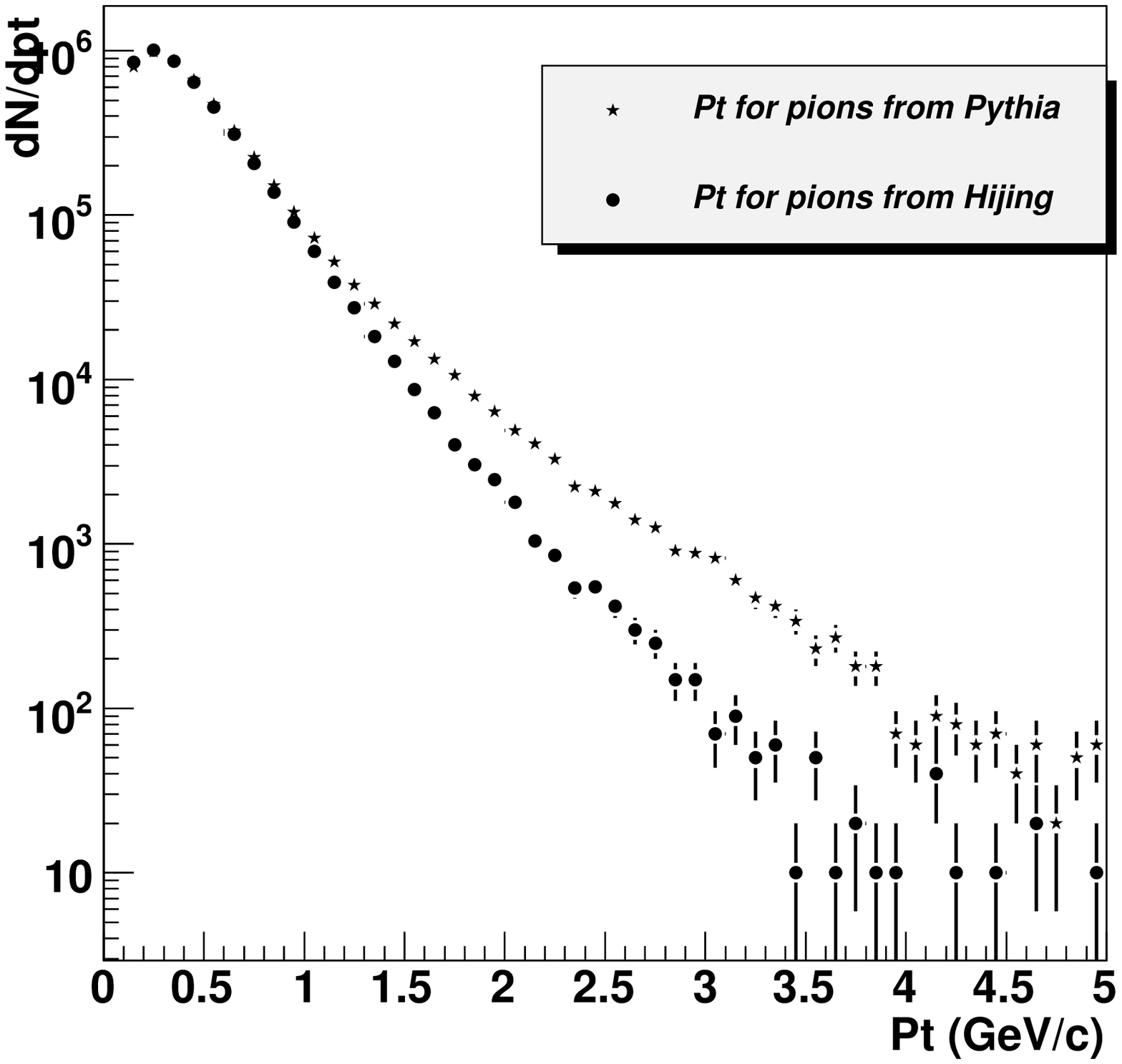}
\includegraphics[height=.3\textheight]{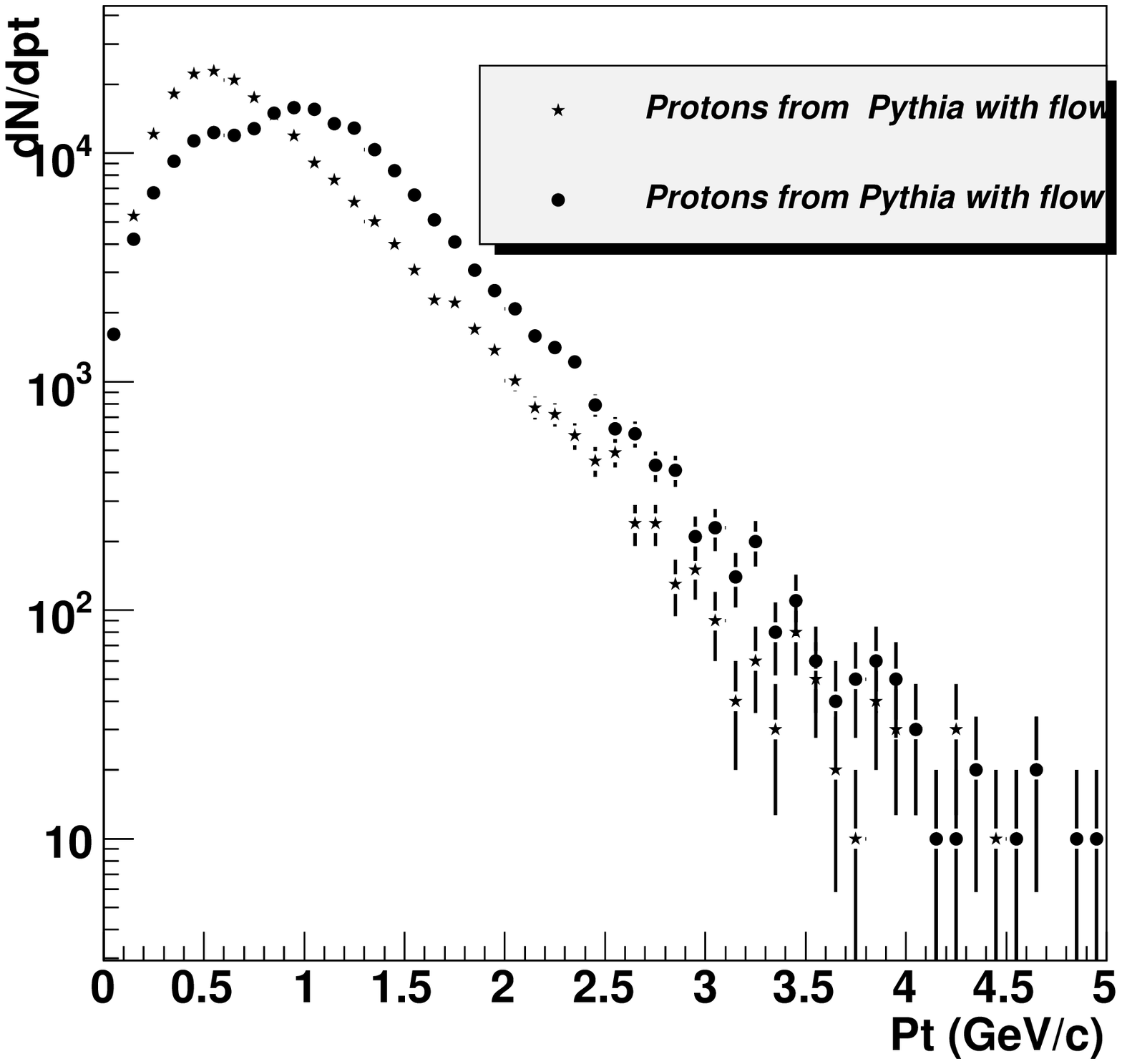}
\caption{Spectra of pions from Pythia and Hijing event generators. The
flow effects are show in the right side, for the protons cases.}
\label{ptHijing-pythia}
\end{figure}

In the left side of Fig.~\ref{ptHijing-pythia} we show the midrapidity ($|\eta|<1$) transverse momentum
spectra obtained with Pythia and Hijing. It is apparent that the two
generators do not yield for $pp$ collisions the same result. Hijing
predicts a slightly steeper $p_t$ dependence than Pythia. In the right side of the figure~\ref{ptHijing-pythia} we show the effect of the afterburner
onto the flow free spectra for the proton case. We have used here a flow velocity of 0.6c
to be able to clearly demonstrate the effect. As expected the flow free
and afterburner spectra coincide at higher $p_t$.

\begin{figure}
\includegraphics[height=.35\textheight]{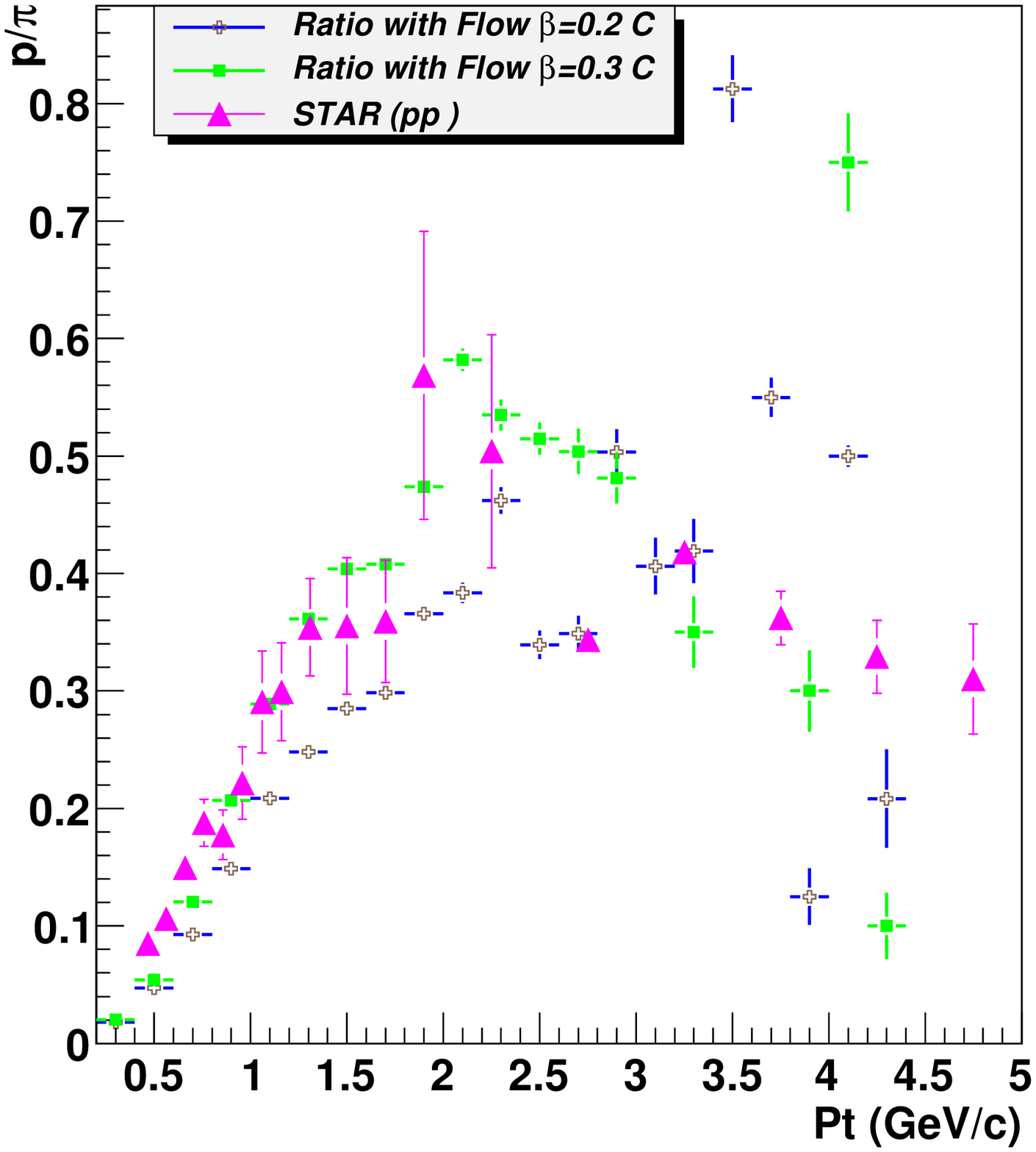}
\includegraphics[height=.35\textheight]{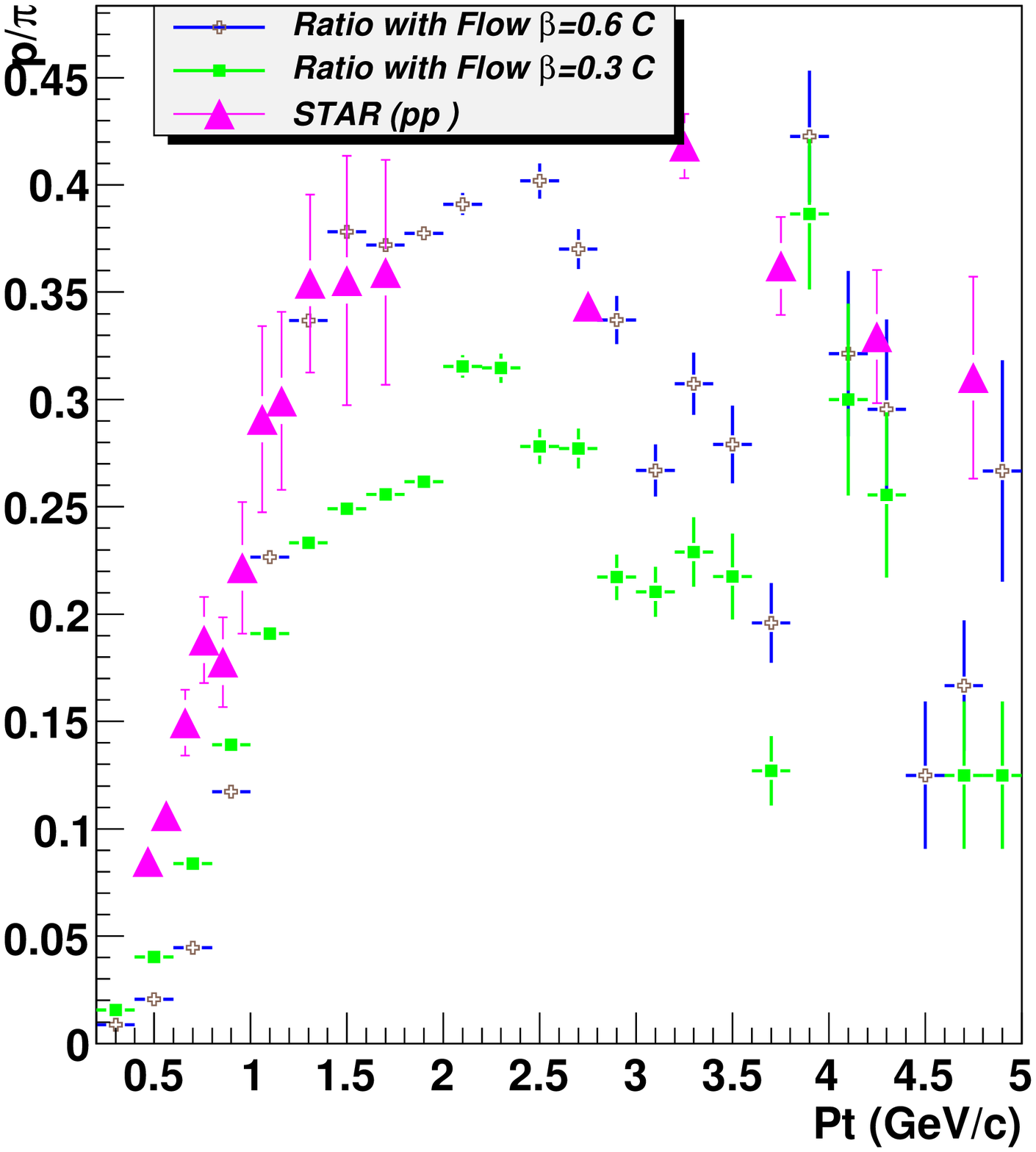}
\caption{Proton  to  pion  ratio  as  function  of  $p_t$.   The  left
correspond  to  experimental results  on  $pp$  collisions from  STAR,
comparing  with our  model using  Pythia. The  right  part shows  our model
using Hijing and comparing it  with the experimental results from $pp$
collisions.}
\label{ppi}
\end{figure}

The ratios p/$\pi$ are shown  in Fig. ~\ref{ppi}.In the left  we show the
results obtained for a flow velocity $\beta$=.2 and .3, applied to the
HIJING events, compared withe STAR pp  data the right side of the same
figure shows  the same data fitted  with Pythia applying  the same flow
parameter and an additional one of 0.6c.  It is obvious that the resulting
fit is  different, Pythia requiring  a much larger flow  velocity than
Hijing to fit the same spectra.

\begin{figure}
\includegraphics[height=.35\textheight]{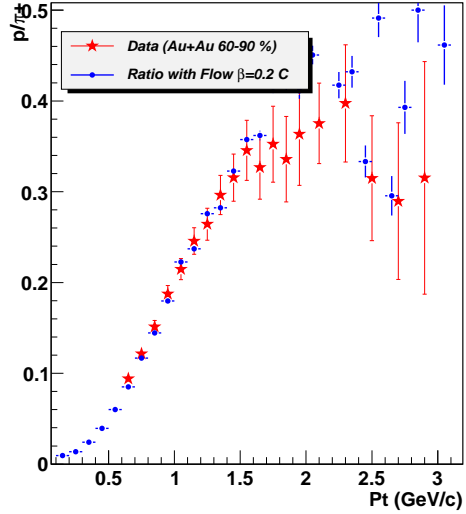}
\caption{Ratio proton to pion from Hijing included radial flow with $\beta =0.2$c and compared with  PHENIX $Au+Au$ peripheral results.}
\label{ptHijing-data}
\end{figure}

The model works well also for peripheral heavy ion collisions as shown in Fig.~\ref{ptHijing-data} comparing with PHENIX~\cite{phenix} data.

\subsection{SUMMARY}
The data of pp collisions at  RHIC suggest that even in pp collisions
the  transverse  momentum  spectra  of  different  particles  are  not
completely parallel to each other.  The analysis suggests that a {\it
flow} velocity albeit  small has to be added.  Without entering in the
foundation of the  existence or not of flow we  have constructed a toy
model where we apply a given  quantity of flow to flow free transverse
momentum spectra generated by either  Pythia with the popcorn
baryon  production mechanism or Hijing. The obtained  results  demonstrate that
both for $pp$ collisions and peripheral heavy ion collisions we obtain a
remarkably  good  fit  to  the  data  using  Hijing  spectra  and  the
velocities  extracted  in the  experiments.   The  differences in  the
quality of  the fit  observed using Pythia  indicate that  the initial
shape of  the spectrum is key to  a good reproduction of  the data. We
conclude that  the flow may  be one of  the key ingredients in  the so
called Baryon puzzle at RHIC.


\begin{theacknowledgments}
We  would like to  thanks A.  Morsch for his comments  and suggestions.
This work was supported in  part by project IN107105 and Conacyt under
grant number 40025-F.

\end{theacknowledgments}


\bibliographystyle{aipproc}   

\end{document}